\documentclass[10pt, preprint2]{aastex}

\newcommand{\ka}{EPIC 202900527} 
\newcommand{\kb}{EPIC 206155547}
\newcommand{\kc}{EPIC 206432863}
\newcommand{\kab}{EPIC 202900527~B} 
\newcommand{\kbb}{EPIC 206155547~B}
\newcommand{\kcb}{EPIC 206432863~B}

\newcommand{\Nkeckrv}{20}
\newcommand{\Nkeckrvb}{7}
\newcommand{\Nkeckrvc}{13}
\newcommand{\Nkecknights}{18}
\newcommand{\Ntresrv}{11}

\newcommand{\rsun}{\ensuremath{R_\sun}}
\newcommand{\msun}{\ensuremath{M_\sun}}

\newcommand{\rjup}{\ensuremath{R_{\rm J}}}
\newcommand{\mjup}{\ensuremath{M_{\rm J}}}

\newcommand{\teff}{\ensuremath{T_{\rm eff}}}

\newcommand{\logg}{\ensuremath{\log g}}
\newcommand{\feh}{[Fe/H]}
\newcommand{\vsini}{\ensuremath{V\sin(I)}}

\newcommand{\ecosw}{\ensuremath{\sqrt{e}\cos\omega}}
\newcommand{\esinw}{\ensuremath{\sqrt{e}\sin\omega}}

\newcommand{\kms}{km~s$^{-1}$}
\newcommand{\ergscm}{erg s$^{-1}$ cm$^{-2}$}

\newcommand{\ik}{{\it Kepler}}
\newcommand{\kt}{{\it K2}}

\newcommand{\sig}[1]{\ensuremath{#1\sigma}}
\newcommand{\figr}[1]{Fig.~\ref{fig:#1}}
\newcommand{\secr}[1]{Sec.~\ref{sec:#1}}

\newcommand{\tabr}[1]{\mbox{Table~\ref{tab:#1}}}

\shorttitle{Three misclassified exoplanets}
\shortauthors{Shporer et al.}

\begin{document}

\title{Three statistically validated \kt\ transiting warm Jupiter exoplanets confirmed as low-mass stars}

\author{
Avi Shporer\altaffilmark{1}, 
George Zhou\altaffilmark{2},
Andrew Vanderburg\altaffilmark{2,7},
Benjamin J.~Fulton\altaffilmark{1},
Howard Isaacson\altaffilmark{3},
Allyson Bieryla\altaffilmark{2},
Guillermo Torres\altaffilmark{2},
Timothy D.~Morton\altaffilmark{4},
Joao Bento\altaffilmark{5},
Perry Berlind\altaffilmark{2},
Michael L.~Calkins\altaffilmark{2},
Gilbert A.~Esquerdo\altaffilmark{2},
Andrew W.~Howard\altaffilmark{6},
David W.~Latham\altaffilmark{2}
}

\altaffiltext{1}{Division of Geological and Planetary Sciences, California Institute of Technology, Pasadena, CA 91125, USA}
\altaffiltext{2}{Harvard-Smithsonian Center for Astrophysics, 60 Garden Street, Cambridge, MA 02138, USA}
\altaffiltext{3}{Astronomy Department, University of California, Berkeley, CA, USA}
\altaffiltext{4}{Department of Astrophysical Sciences, Princeton University, Princeton, NJ 08544, USA}
\altaffiltext{5}{Research School of Astronomy and Astrophysics, Mount Stromlo Observatory, Australian National University, Weston, ACT 2611, Australia.}
\altaffiltext{6}{Department of Astronomy, California Institute of Technology, Pasadena, CA, USA}
\altaffiltext{7}{National Science Foundation Graduate Research Fellow}

\begin{abstract}

We have identified three \kt\ transiting star-planet systems, K2-51 (EPIC~202900527), K2-67 (EPIC~206155547), and K2-76 (EPIC~206432863), as stellar binaries with low-mass stellar secondaries. The three systems were statistically validated as transiting planets, and through measuring their orbits by radial velocity monitoring we have derived the companion masses to be $0.1459^{+0.0029}_{-0.0032}$ \msun\ (\kab), $0.1612^{+0.0072}_{-0.0067}$ \msun\ (\kbb), and $0.0942 \pm 0.0019$ \msun\ (\kcb). Therefore they are not planets but small stars, part of the small sample of low-mass stars with measured radius and mass. The three systems are at an orbital period range of 12--24 days, and the secondaries have a radius within 0.9--1.9 \rjup, not inconsistent with the properties of warm Jupiter planets. 
These systems illustrate some of the existing challenges in the statistical validation approach. We point out a few possible origins for the initial misclassification of these objects, including poor characterization of the host star, the difficulty in detecting a secondary eclipse in systems on an eccentric orbit, and the difficulty in distinguishing between the smallest stars and gas giant planets as the two populations have indistinguishable radius distributions. Our work emphasizes the need for obtaining medium-precision radial velocity measurements to distinguish between companions that are small stars, brown dwarfs, and gas giant planets.

\end{abstract}


\section{Introduction}
\label{sec:intro}

Space-based surveys (\ik, \citealt{borucki16}; \kt, \citealt{howell14}) are producing an increasing number of transiting planet candidates \citep[e.g.,][]{coughlin16, crossfield16, vanderburg16}. Those candidates need to be examined by gathering additional data, to check whether the transit light curve is produced by a transiting star-planet system, or, by a different scenario, making the object a false positive \citep[e.g.,][]{torres11, fressin13}. As there are insufficient observational resources needed for gathering the amount of data required to investigate the true nature of each transiting planet candidate, and because some planets cannot be confirmed with current observational capabilities, a \emph{statistical validation} approach was developed \citep[e.g.,][]{torres11, torres15, morton16}. This approach uses a relatively small amount of observational follow-up data, typically including a single spectrum and a single high angular resolution image of the target, and is based on estimating the probability that the transit light curve is produced by a transiting star-planet system and not a false positive scenario \citep[e.g.,][]{torres11, morton12}.

Therefore instead of the traditional approach of confirming a transiting planet candidate by measuring its orbit and deriving its mass, the validation approach puts an upper limit on the probability the candidate is a false positive. That upper limit is typically at the 1\% or 0.1\% level in order to declare a candidate as a validated planet \citep[e.g.,][]{montet15, crossfield16, morton16}. Hence the validated planets have measured orbital periods and radii, but their masses are unknown.

As part of a campaign to determine masses of transiting warm Jupiter planets --- gas giant planets receiving stellar irradiation below about 10$^8$ \ergscm, equivalent to orbits beyond about 10 days around Sun-like stars \citep{shporer17} --- we have measured the masses of three of the \kt\ validated planets. The resulting masses are in the range of $0.09-0.16\ \msun$, therefore they are not planets but small stars. Those systems are \ka\ (K2-51), \kb\ (K2-67), and \kc\ (K2-76), validated by \citet{crossfield16}. 

We describe our \kt\ data processing and gathering of spectroscopic data in \secr{obs}. The data analysis is described in \secr{anal}. In \secr{dis} we discuss our findings and briefly explore possible reasons for the misclassification of these stellar binaries as planetary systems. Throughout the text we refer to the transit interchangeably as the primary eclipse. 
Although the three systems have \kt\ numbers (e.g.~K2-51) we refer to them hereafter by their EPIC ID number (e.g.~EPIC~202900527) since the former is reserved for planetary systems. In addition, we refer to the low-mass secondary in each system using the upper case `B' (e.g.~EPIC~202900527~B) since it is a stellar object.

\section{Observations}
\label{sec:obs}

\subsection{\kt}
\label{sec:k2}

The three targets were observed by \kt\ during Campaign~2 (\ka) and Campaign~3 (\kb\ and \kc), in long cadence (29.4 minutes integration time). We reduced the \kt\ light curves following \cite{vanderburg14} and \cite{vanderburg16}. Upon identifying the transits we re-processed the light curves by simultaneously fitting for the transits, \kt\ thruster systematics, and low-frequency variations as described by \cite{vanderburg16}. The phase folded light curves are plotted in \figr{lc}.

\subsection{Keck/HIRES}
\label{sec:hires}

The Keck/HIRES data analyzed and presented here include \Nkeckrv\ spectra at a resolution of R$\sim$60,000. We obtained \Nkeckrvb\ spectra of \kb\ and \Nkeckrvc\ of \kc, during \Nkecknights\ nights from August~1~2015 UT to June~28~2017 UT. 
We have also obtained a Keck/HIRES spectrum of \ka, used only for spectroscopic characterization of the primary star and not for radial velocity (RV) measurement (see \secr{tres} and \secr{spec}).

We used the Keck/HIRES instrumental setup of the California Planet Search \citep{howard09}. Since we can tolerate a medium RV precision, of $\sim$0.1 \kms, we used the so-called telluric lines method where the iodine cell is removed from the light path \citep[see e.g.][their Section 2.2]{shporer16}. Briefly, a wavelength solution is obtained through a nightly exposure of a Thorium-Argon lamp and the RVs are derived by measuring the offset in the position of the telluric absorption bands in the target spectra and that of a reference B-type star \citep{chubak12}. The RV due to Earth's barycentric motion is then removed, resulting in the target's absolute systemic velocity \citep{nidever02, chubak12}.

We used exposure times of 1.5--20 minutes, depending on target brightness, and the spectra we obtained have a signal-to-noise ratio (SNR) of 20--40 per pixel. Keck/HIRES RV measurements are listed in \tabr{rv}, and the phase folded RV curves of \kb\ and \kc\ are shown in \figr{rv}.

\subsection{FLWO 1.5m/TRES}
\label{sec:tres}

We obtained \Ntresrv\ spectra of \ka\ using the Tillinghast Reflector Echelle Spectrograph \citep[TRES;][]{furesz08} at the Fred Lawrence Whipple Observatory (FLWO) 1.5~m telescope on Mount Hopkins, Arizona. The TRES spectra have a resolution of R$\sim$44,000 and were collected between May~26~2015 UT and June~10~2017 UT. We used exposure times between 22--34 minutes which resulted in a SNR per resolution element of 17–-29. 

We reduced and extracted the TRES spectra as described by \cite{buchhave10}. We derived the RVs by cross correlating each spectrum order by order against the observed spectrum with the highest SNR in the wavelength range of 4520--6280~{\AA}. \ka\ \Ntresrv\ TRES RVs are listed in \tabr{rv} and the phase folded RV curve is shown in \figr{rv}.

The reference (or template) spectrum is at BJD = 2457854.95084 and its RV is listed as 0.0 \kms\ in \tabr{rv}. To allow putting the TRES RVs on an absolute scale we determined the template spectrum absolute RV by cross correlating it with a synthetic spectrum to be $-57.87 \pm 0.10$ \kms. This RV offset is not added to the RVs in \tabr{rv} to avoid inflating their uncertainties.



\begin{table}
\centering
\caption{Radial velocities
\label{tab:rv}
}
\vspace{2mm}
\begin{tabular}{crc}
\hline
\hline
Time & \multicolumn{1}{c}{RV} & RV error \\
BJD & \multicolumn{1}{c}{\kms} & \kms \\
\hline
\vspace{-3mm}\\
\multicolumn{3}{c}{\ka\ -- \it FLWO 1.5m/TRES}\\
2457168.87299  & $-$17.812 &   0.089 \\
2457852.92588  &   3.141   &   0.090  \\
2457853.95594  &   7.154   &   0.065  \\
2457854.95084  &   0.000   &   0.090  \\
2457863.93643  & $-$6.752  &   0.090  \\
2457864.95402  & $-$20.111 &   0.175  \\
2457906.81976  &   2.089   &   0.122  \\
2457907.79475  &  $-$9.658 &   0.074  \\
2457908.80884  & $-$15.654 &   0.062  \\
2457909.80521  & $-$17.658 &   0.138  \\
2457914.78970  & $-$11.337 &   0.182  \\  
\hline 
\vspace{-3mm}\\
\multicolumn{3}{c}{\kb\ -- \it Keck/HIRES}\\
2457354.82966  &  34.953  &   0.276\\
2457652.05475  &  33.550  &   0.258\\
2457654.00564  &  35.045  &   0.367\\
2457887.11446  &  51.393  &   0.231\\
2457907.09022  &  51.378  &   0.637\\
2457908.08529  &  54.720  &   0.108\\
2457909.09961  &  56.853  &   0.040\\
\hline
\vspace{-3mm}\\
\multicolumn{3}{c}{\kc\ -- \it Keck/HIRES}\\
2457236.12417 &  $-$11.776  &   0.115 \\
2457652.97550 &      0.658  &   0.577 \\
2457653.99404 &   $-$1.006  &   0.103 \\
2457678.89215 &   $-$5.313  &   0.051 \\
2457713.86465 &   $-$0.755  &   0.072 \\
2457747.76545 &      0.362  &   0.205 \\
2457760.73066 &      0.791  &   0.095 \\
2457887.10573 &  $-$10.511  &   0.122 \\
2457910.10825 &  $-$13.870  &   0.052 \\
2457913.08178 &   $-$4.266  &   0.244 \\
2457926.10787 &   $-$1.742  &   0.103 \\
2457927.13167 &    0.076    &   0.212 \\
2457933.02457 & $-$16.261   &   0.144 \\
\hline
\hline
\end{tabular}
\end{table}

\begin{figure*}
\includegraphics[scale=0.4]{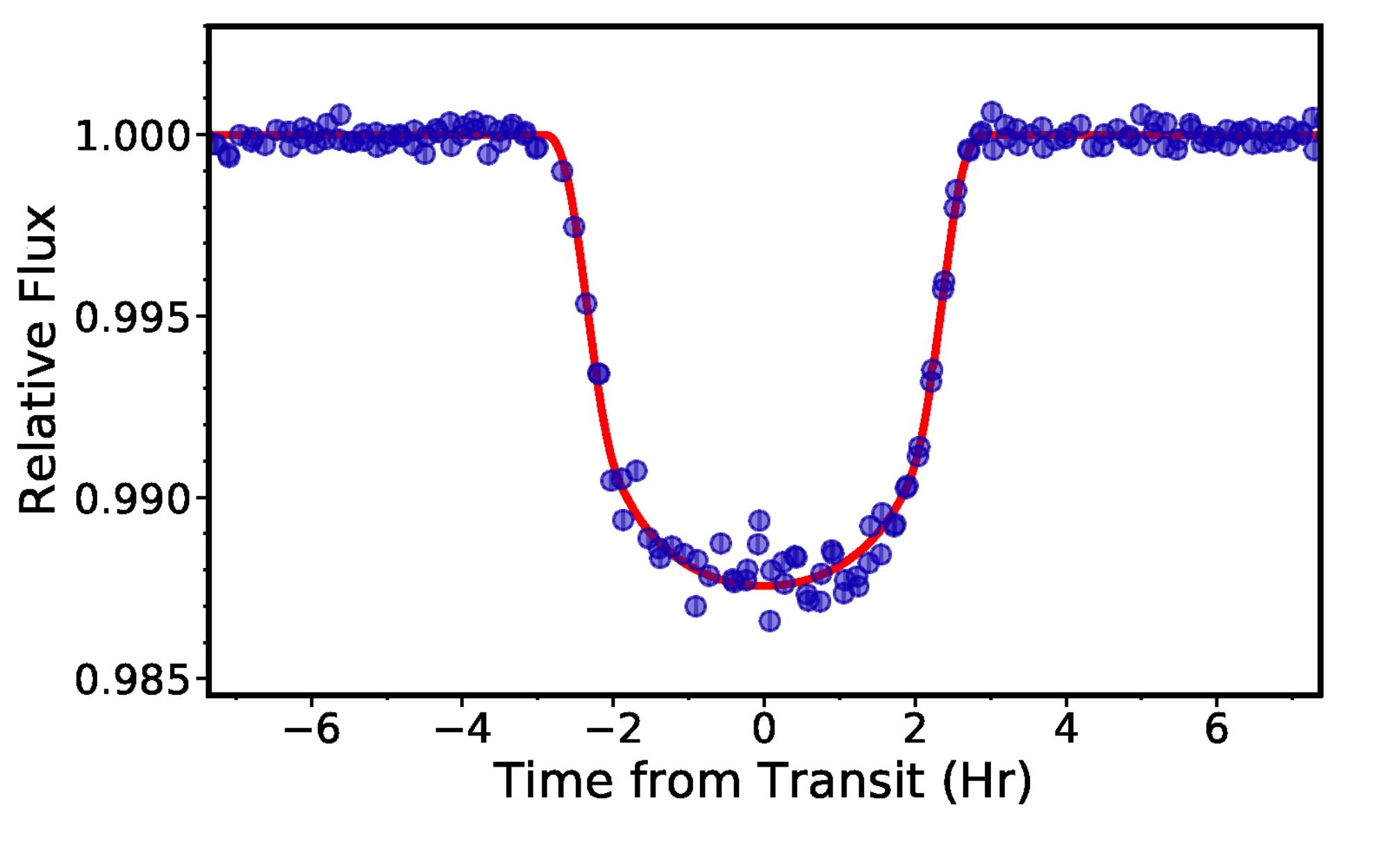}
\includegraphics[scale=0.4]{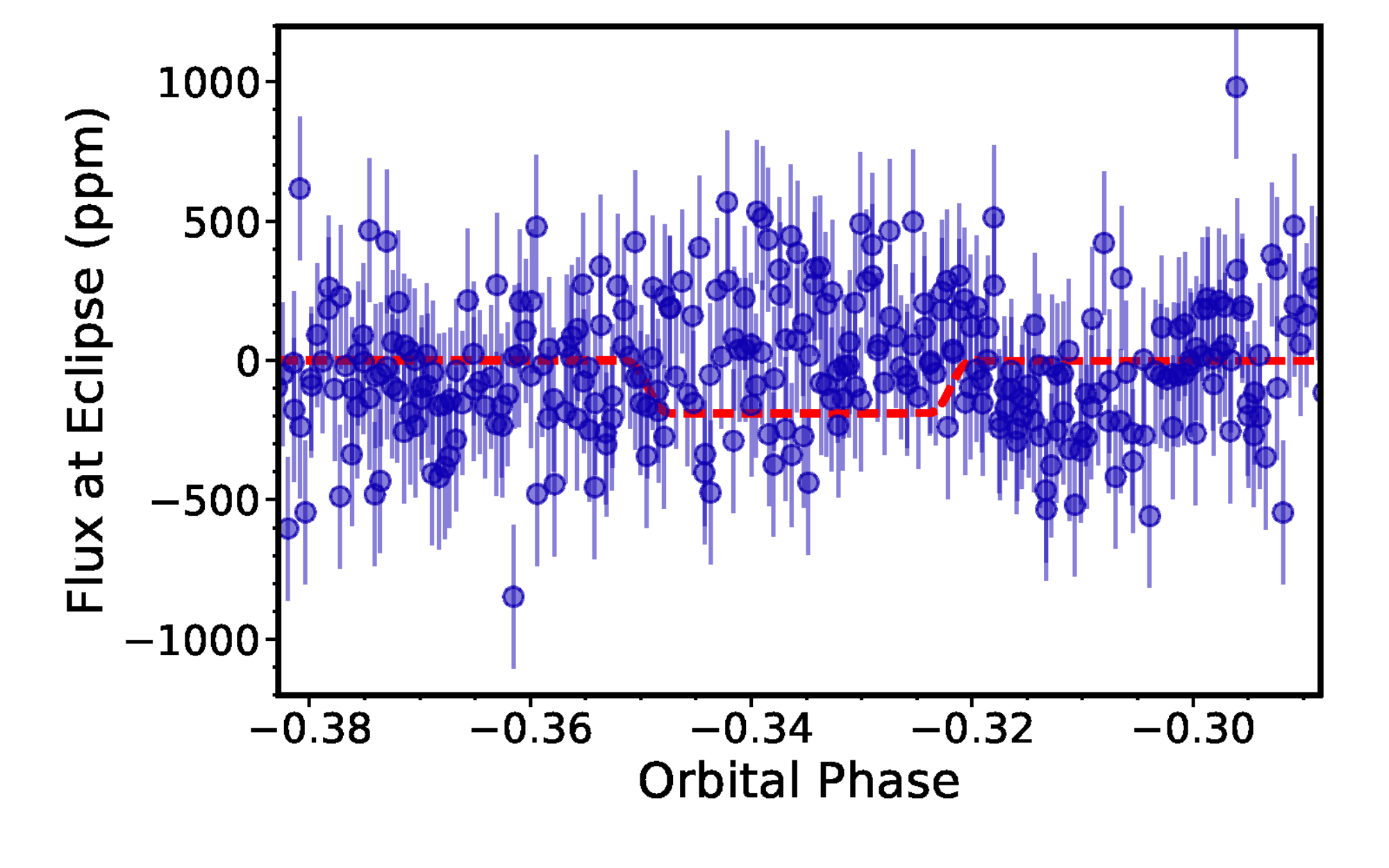}
\includegraphics[scale=0.4]{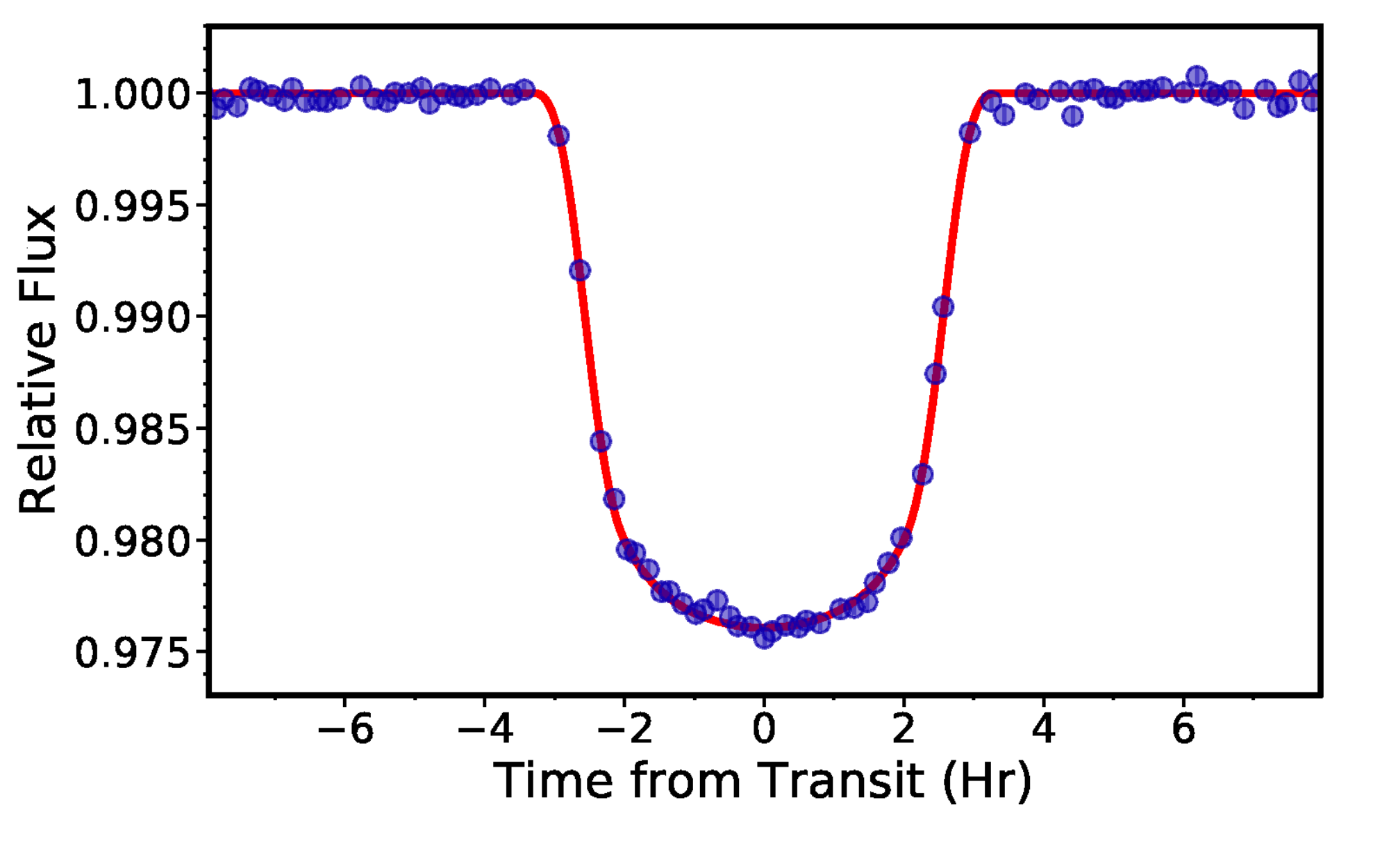}
\includegraphics[scale=0.4]{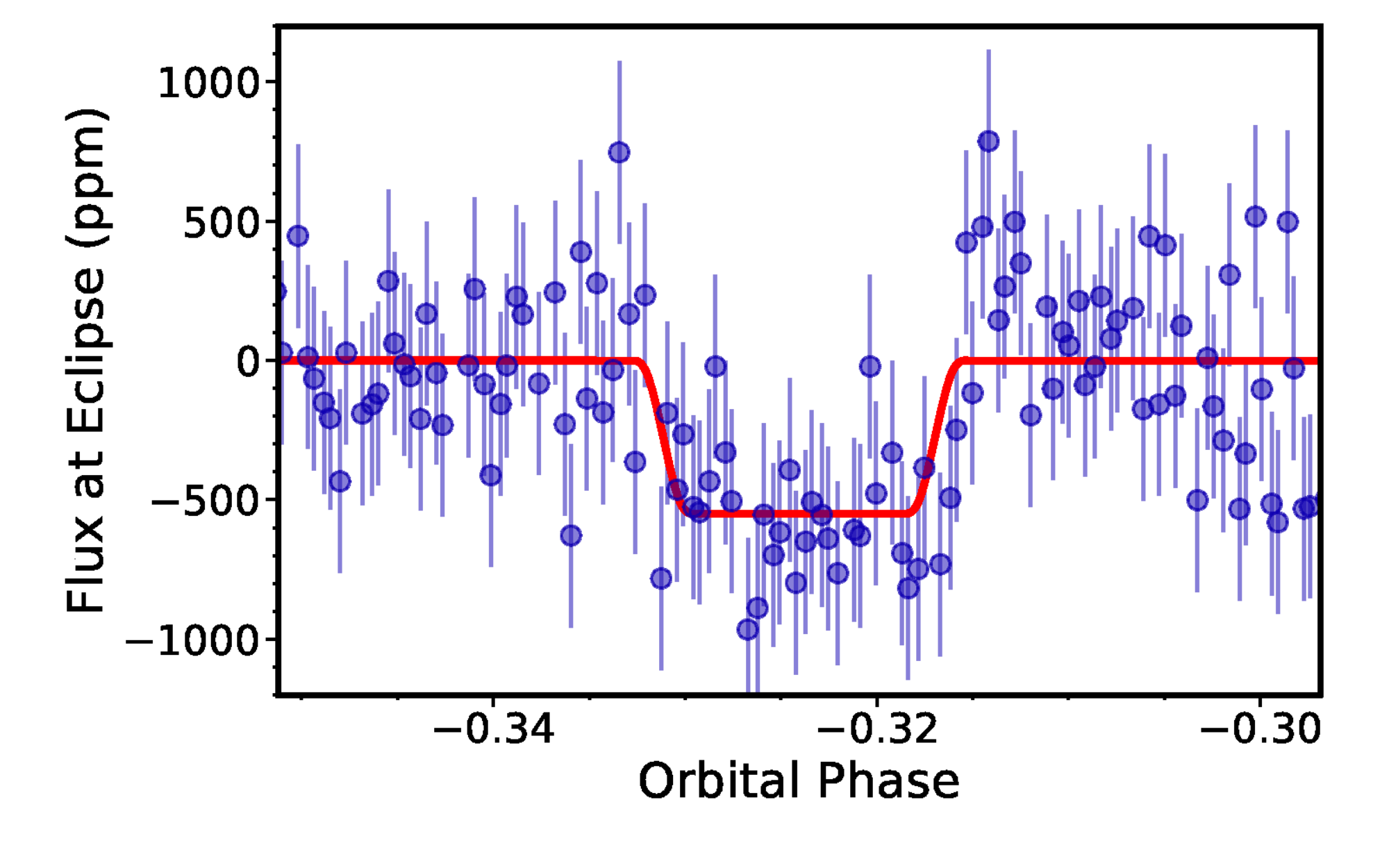}
\includegraphics[scale=0.4]{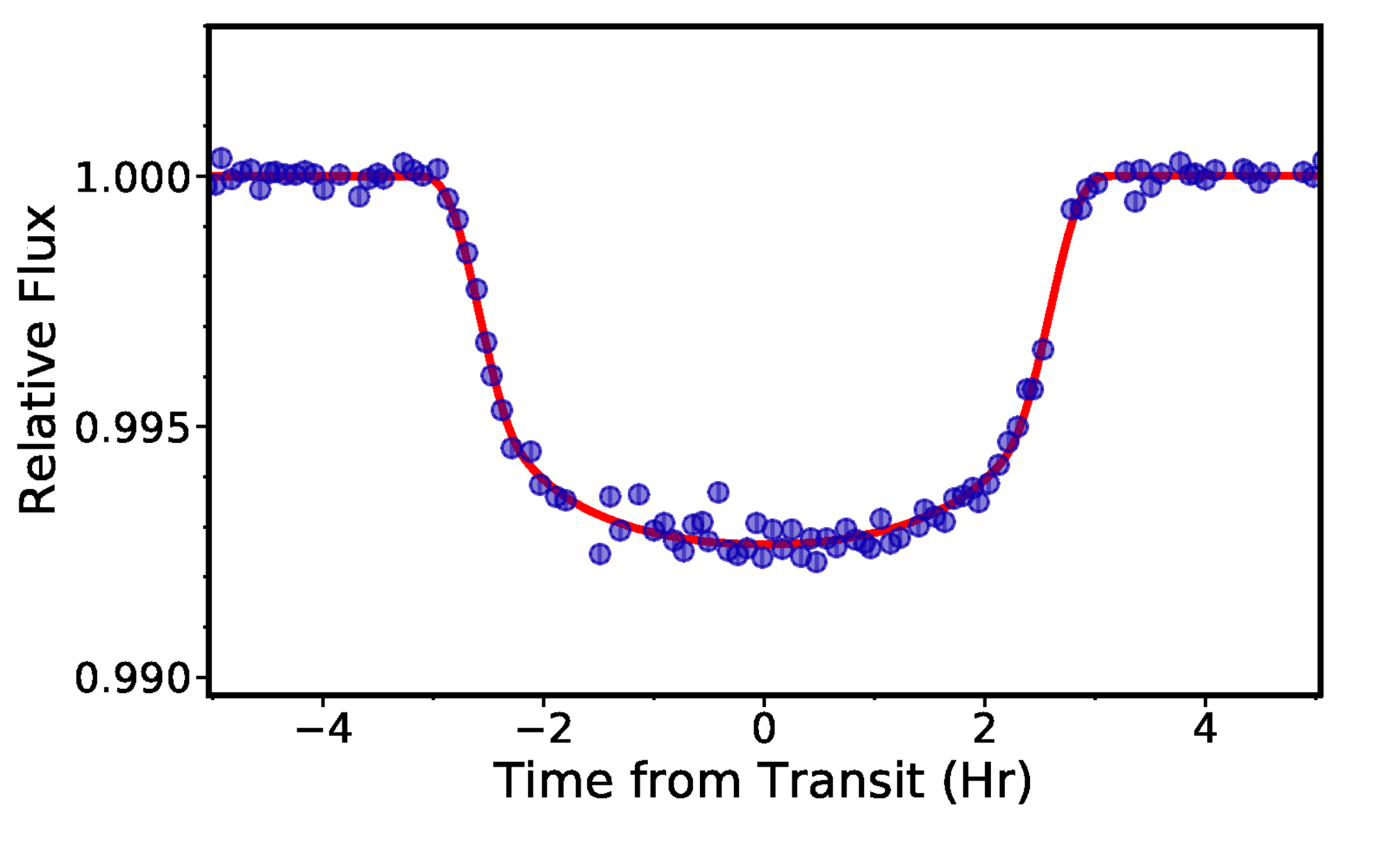}
\includegraphics[scale=0.4]{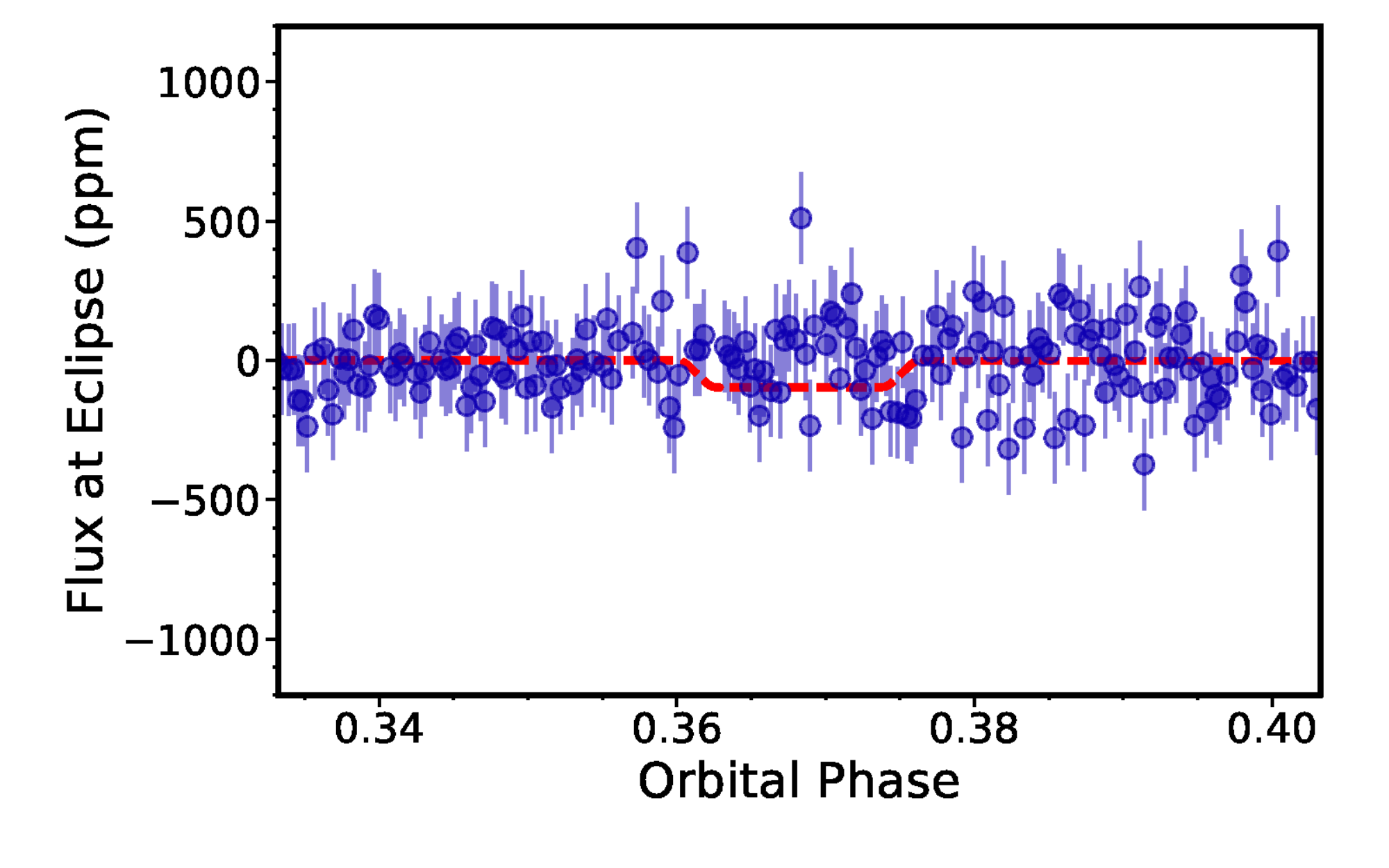}
\caption{Normalized and phase folded \kt\ light curves of the transit (left column) and secondary eclipse (right column; plotted in ppm) of \ka\ (top row), \kb\ (middle row), and \kc\ (bottom row). \kt\ measurements are in blue, and the fitted model is plotted with a solid red line for the three transits and the secondary eclipse of \kb. For \ka\ and \kc\ the plotted secondary eclipse models show the \sig{3} upper limit on the eclipse depth, plotted with a dashed red line. All measurements are plotted with error bars which in the transit light curve panels are smaller than the marker size.
\label{fig:lc}}
\end{figure*}

\begin{figure}
\includegraphics[scale=0.4]{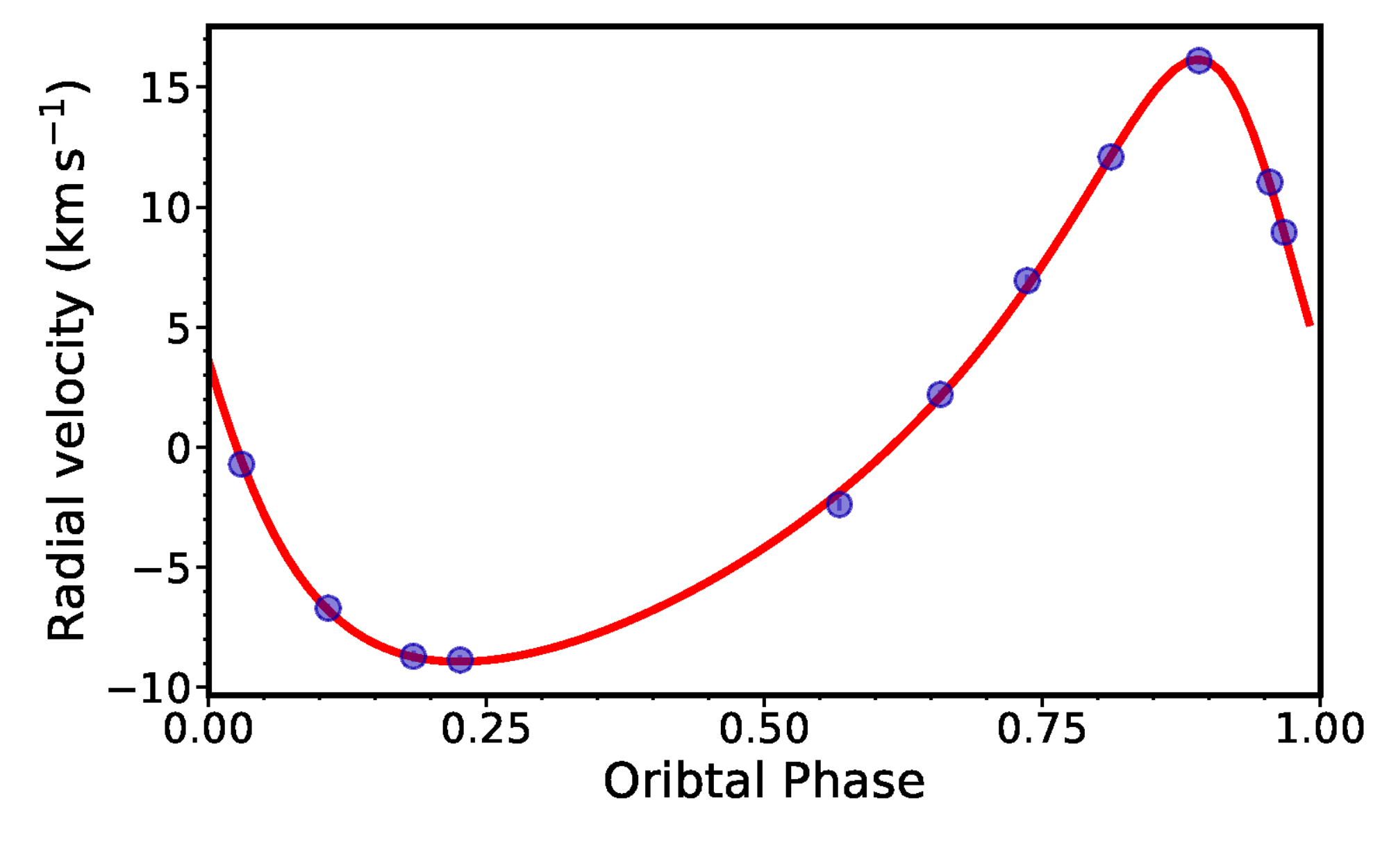}
\includegraphics[scale=0.4]{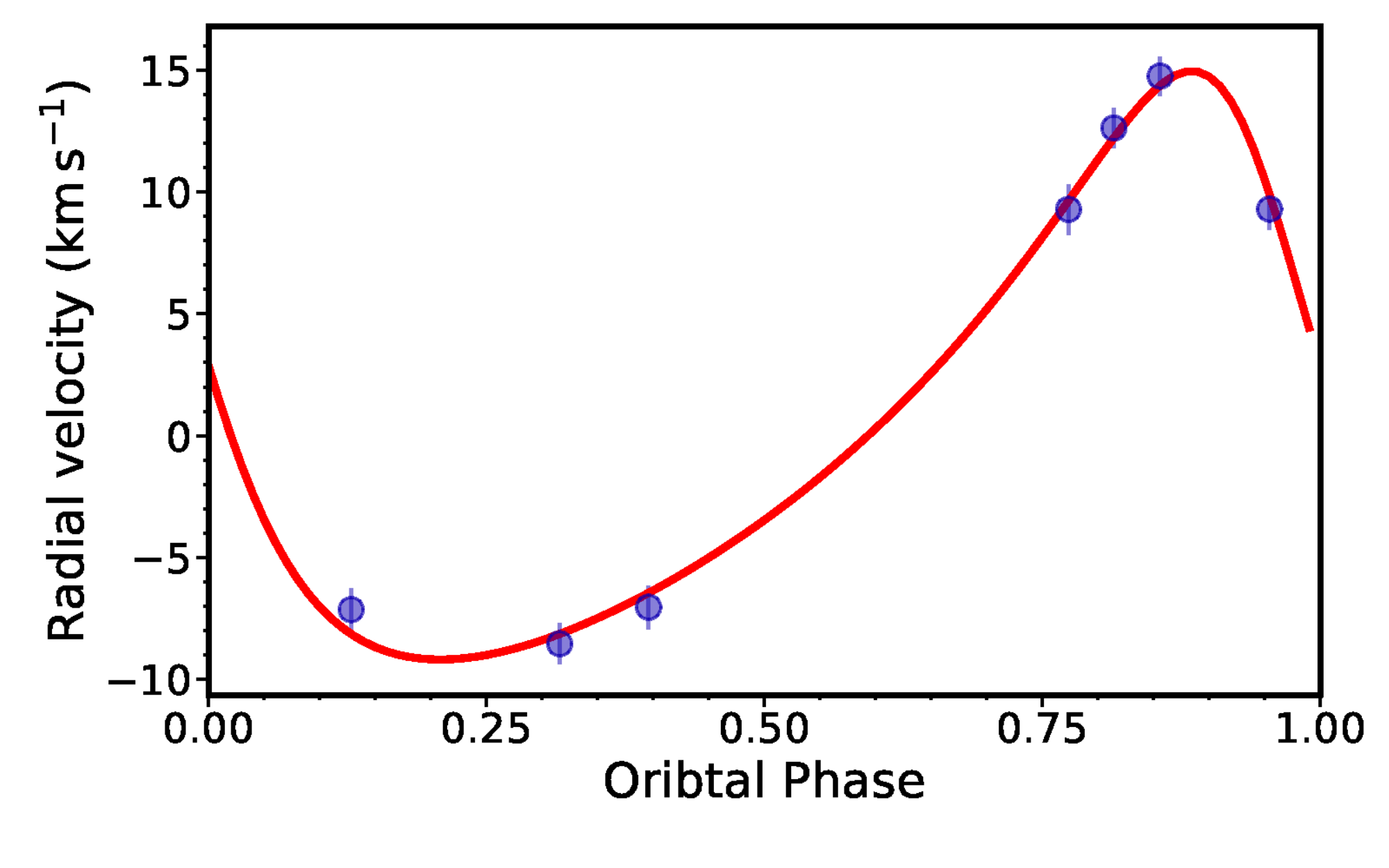}
\includegraphics[scale=0.4]{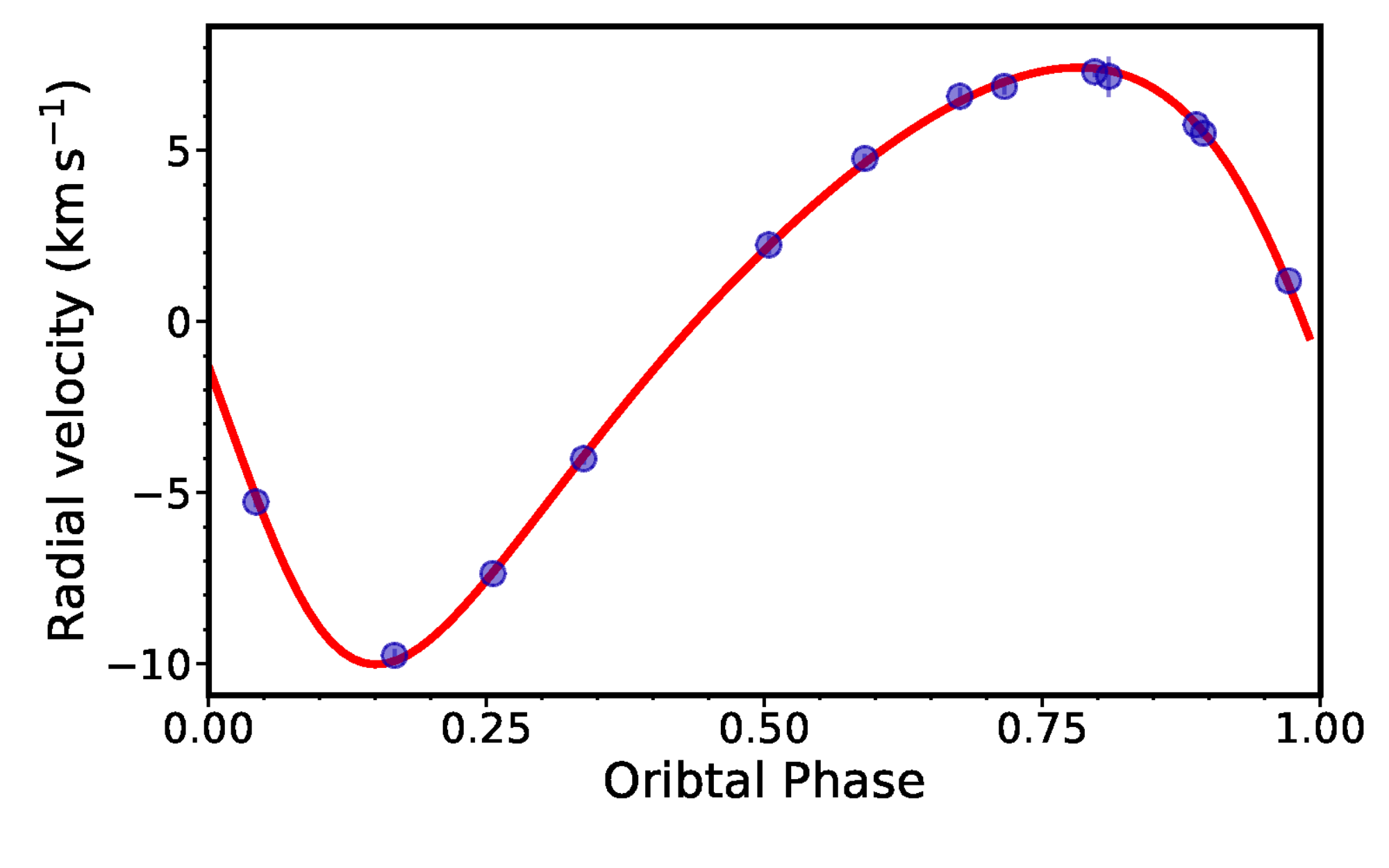}
\caption{Phase folded RV curves, after subtracting the RV zero point $\gamma$, of \ka\ (top panel), \kb\ (middle panel), and \kc\ (bottom panel). The transit is at phase zero. RVs are marked in blue and the fitted orbit models are marked in solid red lines. All measurements are plotted with error bars which are often smaller than the marker size.
\label{fig:rv}}
\end{figure}

\section{Data analysis and results}
\label{sec:anal}

\subsection{Stellar spectroscopic parameters}
\label{sec:spec}

We derived the spectroscopic stellar parameters using the \texttt{SpecMatch} package \citep{petigura15, petigura17} and the iodine-free HIRES spectra of each star. Those include the effective temperature \teff, surface gravity \logg, metallicity \feh, and stellar rotation projected on the line-of-sight \vsini\ where $V$ is the equatorial rotation and $I$ is the stellar rotation inclination angle. We averaged the parameters extracted from all of the individual observations. The observation-to-observation parameter variance was smaller than the quoted \sig{1} uncertainties in all cases. The \texttt{SpecMatch} results are listed in \tabr{params}.

\subsection{Global model fitting}
\label{sec:fit}

We performed a global modeling of the available photometric and RV measurements, along with the spectroscopic atmospheric properties of the primary star, to derive the parameters for each system. For stellar binaries, the transit-derived orbital semi-major axis normalized by the primary stellar radius $a/R_1$ is dependent on the sum of the two components' masses $M_1+M_2$ and the volume of the primary star, as per \citet{sozzetti17}:
\begin{equation}\label{eq:ars}
\left(\frac{a}{R_1}\right)^3  = \frac{G}{4 \pi^2} P^{2} \frac{M_1 + M_2}{R_1^3} \, .
\end{equation}
Where $P$ is the orbital period and $G$ the gravitational constant. To take advantage of this relation we fit directly for the masses of the two stars, the primary radius, and the secondary to primary radii ratio $R_2/R_1$, as well as the standard transit and RV orbital parameters including the orbital period $P$, mid-transit time $T_0$, line-of-sight orbital inclination $i$, orbital eccentricity parameters \ecosw\ and \esinw\ (where $e$ is the orbital eccentricity and $\omega$ the argument of periastron), RV zero point $\gamma$, RV jitter $s$, and the primary metallicity \feh. We include the secondary eclipse in our model where the eclipse depth is the secondary to primary flux ratio in the \ik\ band $(F_2/F_1)$. We used the model of \citet{mandel02} for the transit and secondary eclipse light curves.

At each iteration, we calculate a normalized orbital semi-major axis $a/R_1$ as per Equation~\ref{eq:ars}, and an orbital RV semi-amplitude $K$ from the masses and eccentricities tested. To constrain the stellar masses and radii we interpolate the Dartmouth isochrones \citep{dotter08} at each step over the parameters $M_1$, $R_1$, and \feh, to derive an expected \teff\ value. 
We then compare the isochrone-derived \teff\ with that measured spectroscopically and add the difference as a penalty term to the likelihood function.
We apply a similar penalty in the likelihood function for the primary star's \logg\ value, calculated from the tested $M_1$ and $R_1$ values, by comparing it to that measured spectroscopically. 
The stellar metallicity \feh\ is constrained by a Gaussian prior over its spectroscopically measured value. The remaining parameters are assumed to have uniform priors. The RV jitter $s$ is calculated as per \citet{haywood16}. 

Quadratic limb darkening coefficients, $u_1$ and $u_2$, are interpolated from \citet{claret04} to the atmospheric parameters of each star, and held fixed during the fitting process.

We explore the posterior probability distributions via a Markov chain Monte Carlo analysis, using the affine invariant ensemble sampler \emph{emcee} \citep{emcee13}. The 68.3\% confidence regions for the MCMC free parameters, as well as several inferred parameters, are listed in \tabr{params}. The inferred parameters include, in addition to parameters mentioned above, \ik\ band luminosity of the primary $L_1$, and the secondary $L_2$, system age, impact parameter of the transit $b$, and of the secondary eclipse (occultation) $b_\mathrm{occ}$, transit duration $T_{14}$, ingress duration $T_{12}$, and secondary eclipse phase where primary eclipse (transit) phase is taken as phase zero. The best fit transit and secondary eclipse light curve models are shown in \figr{lc} and the orbital RV curve models in \figr{rv}.

For the most part the transit parameters we derive are similar to those reported by \cite{crossfield16}. One notable exception is the orbital period of \kc. We find that the true orbital period is exactly half the one reported by those authors. Another difference is our detection of a secondary eclipse for \kb, at a depth of $560^{+160}_{-180}$ ppm. For \ka\ and \kc\ we find no detectable secondary eclipses, and place \sig{3} upper limits on their depths of 190 and 97 ppm respectively.

\section{Discussion}
\label{sec:dis}

\begin{figure}[ht]
\hspace{-5mm}
\includegraphics[scale=0.4]{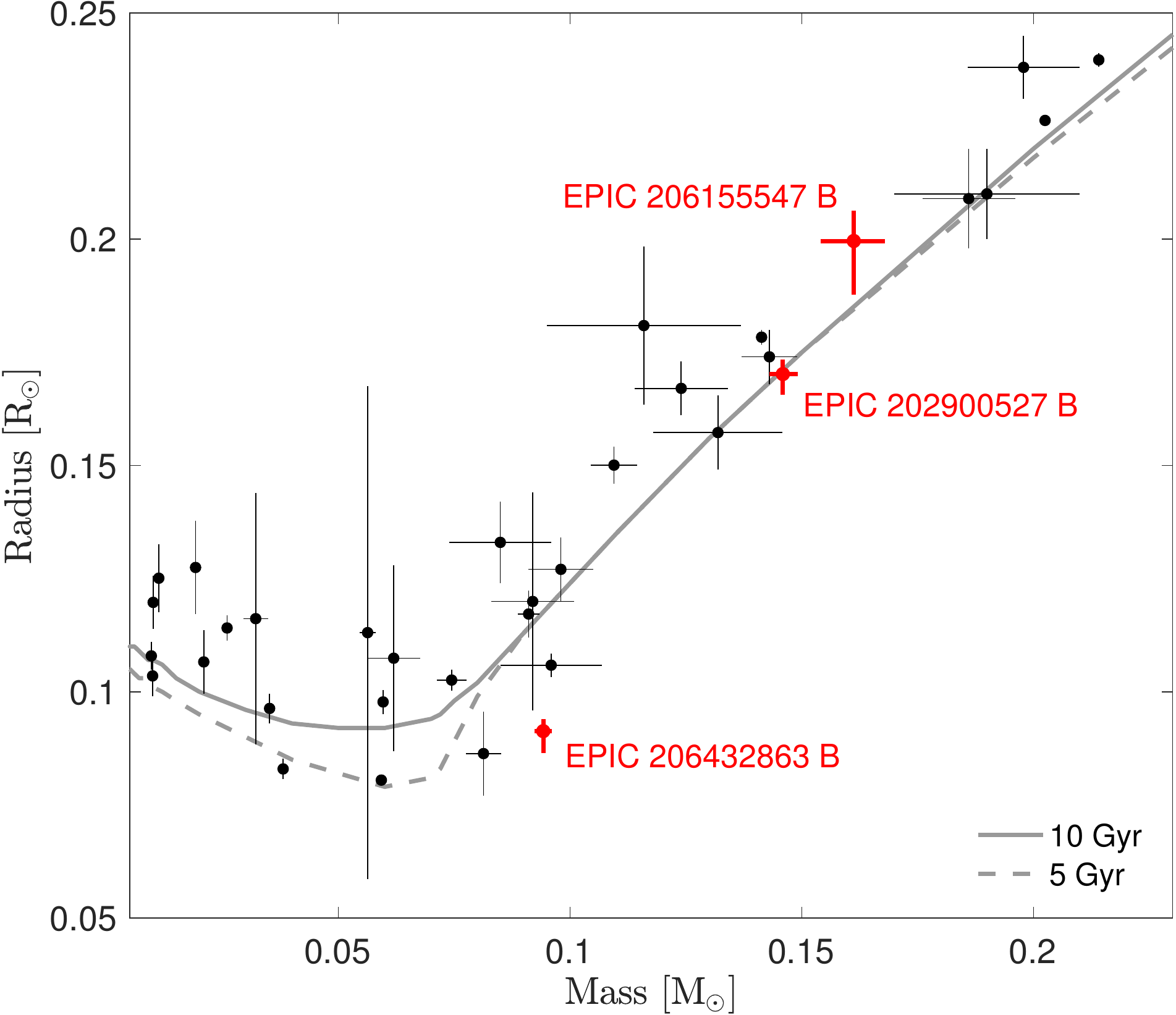}
\caption{Radius-mass diagram for  massive planets, brown dwarfs, and small stars. The three objects studied here are marked in red. The solid and dashed lines are theoretical radius-mass relations for 10 Gyr (solid line) and 5 Gyr (dashed line) old stars \citep{baraffe15} and substellar objects \citep{baraffe03} with solar metallicity.
\label{fig:radmass}}
\end{figure}

The three objects studied here, \kab, \kbb, and \kcb, are among the smallest stars with measured radius and mass. Strictly speaking, the values are model-dependent to some extent as they rely on masses and radii for the primary stars inferred from stellar evolution models. The uncertainties in the host star properties dominate the error budget for the secondaries. The positions of the B components in the radius-mass diagram are shown in \figr{radmass}, compared to other objects ranging from massive planets to brown dwarfs and small stars \citep{pont05, pont06, talor11, southworth11, ofir12, akeson13, diaz13, moutou13, triaud13, diaz14, zhou14, bayliss17, vonboetticher17}. The overplotted lines are theoretical solar metallicity radius-mass relations \citep{baraffe03, baraffe15}. While \kab\ agrees well with the theoretical prediction, \kbb\ appears to be larger than predicted for its mass, and \kcb\ appears to be smaller than predicted for its mass. In fact, \kcb\ is one of the smallest objects with mass just above the theoretical minimum mass required for hydrogen burning, where the behavior of the radius changes from slowly \emph{decreasing} with increasing mass for massive brown dwarfs, to \emph{increasing} with increasing mass for low-mass stars.

\vspace{1.5mm}

As far as we are aware there are no publicly available model predictions over the full mass range shown in \figr{radmass} for metallicities other than solar.  However, the discrepancy for \kcb\ does not appear to be due to metallicity, as our spectroscopic analysis indicates it has \feh $= +0.01 \pm 0.04$, essentially matching the metallicity of the models shown in the figure. Similarly, metallicity is unlikely to explain the inflated radius of \kbb, as this system is metal-poor and available model predictions over this mass range indicate that if anything the effect should go in the opposite direction, making the star smaller \citep[e.g.][]{borrows11}. In principle stellar activity remains a possible explanation for the larger size, as has been found to be the case in many other low-mass binaries \citep[e.g.,][]{torres13}, though in most of those examples the activity is maintained at a high level by tidally induced rapid rotation in short period orbits (typically a few days). This is not expected in \kb\ given its longer period of 24 days and old age ($\approx$10 Gyr). 

The orbits of the three binary systems are moderately eccentric, with precisely measured eccentricities ranging from 0.25 to 0.38. Given their long orbital periods of 12--24 days they are not expected to have been tidally circularized within the stellar lifetime \citep[e.g.,][]{mazeh08}, and the eccentricities are within the range seen in systems with a similar period range (see, e.g., \citealt{mazeh08} Figure 1, \citealt{shporer16} Figure 6).

\subsection{Why were these stellar binaries classified as transiting planets?}
\label{sec:why}

The three systems discussed here were validated as planets by \cite{crossfield16} based on a statistical procedure that considered possible false positive scenarios. That statistical validation procedure results in the relative likelihoods of the transit signals being due to a false positive or a true planet. The reported false positive probabilities (FPPs) were $\sim$10$^{-3}$ for the \ka\ and \kb\ systems and $\sim$10$^{-4}$ for \kc. 
Our identification of these objects as stellar binaries raises the question of why they were initially misclassified as planets.

For \ka\ and \kb\ the stellar companions' radii derived by \cite{crossfield16} are smaller than derived here by 30 \% and 20 \%, respectively. This is clearly one of the primary reasons for the low FPP estimated by \cite{crossfield16} for these two systems. The smaller companions' radii followed from a smaller estimate of the host star radius by 20 \% for both systems, and a measured secondary-to-primary radii ratio which is 15 \% smaller for \ka.

To investigate the origin of the smaller host star radii derived by \cite{crossfield16} we looked into their stellar characterization calculations (I.~Crossfield, private communication) using the {\it isochrones} package \citep{morton15a}, which are then used by the {\it vespa} package \citep{morton15b} to calculate the FPPs. These calculations use optical spectra and broad band photometry, when available, from APASS, 2MASS, and WISE. We noticed that in these two cases the fitted stellar model is a poor fit to the data and is inconsistent with at least some of the input measurements. We believe this resulted from poor quality broad band photometric measurements with underestimated uncertainties. Therefore we conclude it is the host stars' poor characterization which led to the underestimated companion radii and the underestimated FPPs. Similar cases of poor stellar characterization can be identified by visually examining the {\it isochrone} output diagnostic plots, or by calculating a goodness-of-fit metric.

The success of validation methods often relies on the ability to rule out the presence of secondary eclipses, which for systems with eccentric orbits, as those studied here, does not necessarily occur half an orbit away from the transit. While the procedure of \cite{crossfield16} did include a search for secondary eclipses throughout the entire orbital phase, it assumed an eclipse duration equal to that of the transit, which is usually not the case in eccentric systems. For \ka, \kb, and \kc\ the expected secondary eclipse durations are a factor of 1.66, 1.69, and 0.67 times the primary eclipse duration, respectively. As described in \secr{fit} we have searched for secondary eclipses as part of the global modeling. While we do not detect a secondary eclipse for \ka\ and \kc, we do detect an eclipse for \kb, the largest and most massive of the three objects, at close to \sig{3} significance. The measured eclipse depth of $560^{+160}_{-180}$ ppm is consistent with a stellar secondary, and is at least an order of magnitude larger than the expected depth in case the secondary is substellar. The nondetection of this secondary eclipse by \citet{crossfield16} might be related to the fact that it is 1.69 times longer than the transit. 
Although as noted earlier the misclassification of \kb\ resulted from poor host star characterization, a detection of the secondary eclipse would have immediately led to classifying the companion as stellar.

We note that the radius distributions of large planets and small stars overlap, making it difficult for validation procedures to distinguish between the two kinds of objects. 
For \kcb\ this becomes especially difficult, since its mass of 0.0942 $\pm$ 0.0019 \msun\ (= 98.7 $\pm$ 2.0 \mjup) is close to the theoretical minimal stellar mass required for burning hydrogen, and, its radius of 0.0913$^{+0.0048}_{-0.0026}$ \rsun\ (= 0.888$^{+0.047}_{-0.025}$ \rjup) is fully consistent with radii of non-inflated planets. As shown in \figr{radmass} it is smaller than theoretically expected for its mass, making it further difficult to be identified as a stellar object through statistical validation. In addition, for \kcb\ \cite{crossfield16} report an orbital period that is twice the true value, which may have also affected the validation calculations.

\section{Summary and conclusions}
\label{sec:sum}

We have identified three of the \kt\ statistically validated warm Jupiter planets to be stellar binary systems with low-mass secondaries. 
We presented a few possible explanations for the misclassification, including poor host star characterization, 
the difficulty in identifying shallow secondary eclipses of long period eccentric systems, 
and the difficulty to distinguish between small stars and gas giant planets. 
As a whole the misclassification of the three systems identified here presents the existing challenges in the validation approach especially when applied to long period systems and/or gas giant planet candidates. 
Their correct classification, shown here (along with three other validated planets identified by \citealt{cabrera17} as false positives), makes them good test cases for further improvement of statistical validation techniques of transiting planet candidates, which in turn will support current and future transiting planet surveys including \kt, TESS \citep{sullivan15}, and PLATO \citep{rauer14}.

The medium-precision RVs we have obtained here, with a precision at the level of 0.1~\kms, exemplifies their efficiency in identifying transiting planet candidates where the transiting object is a low-mass star or a brown dwarf.

The three objects studied here belong to the small sample of low-mass stars with measured mass and radius. Further extending that sample will lead to better understanding of small stars and the processes shaping their radius-mass relation.

\acknowledgments

A.~V.~is supported by the NSF Graduate Research Fellowship, grant No.~DGE 1144152.
G.~T.~acknowledges partial support for this work from NSF grant AST-1509375 and NASA grant NNX14AB83G (\ik\ Participating Scientist Program).
D.~W.~L.~acknowledges partial support from the \ik\ mission via Cooperative Agreement NNX13AB58A with the Smithsonian Astrophysical Observatory.
This paper includes data collected by the \kt\ mission. Funding for the \kt\ mission is provided by the NASA Science Mission directorate.
Some of the data presented herein were obtained at the W.~M.~Keck Observatory, which is operated as a scientific partnership among the California Institute of Technology, the University of California and the National Aeronautics and Space Administration. The Observatory was made possible by the generous financial support of the W.~M.~Keck Foundation.

{\it Facilities:} 
\facility{\kt}, 
\facility{Keck:I (HIRES)},
\facility{FLWO:1.5m (TRES)}



\begin{deluxetable}{lccccccccc}
\tablecaption{Fitted and derived parameters \label{tab:params}}
\tablewidth{0pt}
\tablehead{
\colhead{Parameter} & \multicolumn{3}{c}{\ka\ (K2-51)} & \multicolumn{3}{c}{\kb\ (K2-67)} & \multicolumn{3}{c}{\kc\ (K2-76)}  \\ 
 & \colhead{Value} & \colhead{$+$\sig{1}} & \colhead{$-$\sig{1}} & 
   \colhead{Value} & \colhead{$+$\sig{1}} & \colhead{$-$\sig{1}} &
   \colhead{Value} & \colhead{$+$\sig{1}} & \colhead{$-$\sig{1}}
}
\startdata
\multicolumn{9}{l}{\it Spectroscopic parameters\tablenotemark{a}}\\
\teff\ [K]   & 5548 & 60   & 60   & 5907  & 60   & 60   & 5762 & 60 & 60 \\
\logg\ [cgs] & 4.17 & 0.07 & 0.07 & 4.13  & 0.07 & 0.07 & 4.20 & 0.07 & 0.07\\
\feh  & $+$0.32 & 0.04 & 0.04 & $-$0.32 & 0.04 & 0.04 & $+$0.01 & 0.04 & 0.04\\
\vsini\ [\kms] & 11.4 & 0.5  & 0.5 & $<$ 2 & -- & -- & 5.3 & 1.8 & 1.8 \\
\hline

\multicolumn{9}{l}{\it Fitted parameters\tablenotemark{b}}\\
$P$\ [day]       & 13.00847 & 0.00027 & 0.00018 & 24.38752 & 0.00072 & 0.00067 & 11.98980 & 0.00017 & 0.00018 \\
$T_0-$2456900\ [BJD]     & 5.75715 & 0.00069 & 0.00090 & 85.88408 & 0.00094 & 0.00086 & 83.82617 & 0.00055 & 0.00054 \\
$M_1$\ [\msun] & 1.068 & 0.029 & 0.032 & 0.916 & 0.031 & 0.029 & 0.964 & 0.026 & 0.026 \\
$R_1$\ [\rsun] & 1.695 & 0.049 & 0.037 & 1.399 & 0.079 & 0.056 & 1.171 & 0.060 & 0.033 \\
$M_2$\ [\msun]  & 0.1459 & 0.0029 & 0.0032 & 0.1612 & 0.0072 & 0.0067 & 0.0942 & 0.0019 & 0.0019 \\
$R_2/R_1$     & 0.10047 & 0.00066 & 0.00065 & 0.14261 & 0.00130 & 0.00087 & 0.07843 & 0.00081  & 0.00046 \\
$i$\ [deg]    & 89.98 & 1.08 & 0.97 & 89.37 & 0.43 & 0.52 & 89.35 & 0.43 & 0.42 \\
\ecosw  & 0.403 & 0.010 & 0.016 & 0.452 & 0.016 & 0.017 & $-$0.4081 & 0.0104 & 0.0098 \\
\esinw  & 0.4656 & 0.0078 & 0.0106 & 0.397 & 0.040 & 0.044 & $-$0.2971 & 0.0104 & 0.0098 \\
$\gamma$\tablenotemark{c}\ [\kms]  & -8.945 & 0.082 & 0.081 & 42.09 & 0.36 & 0.36 & -6.506 & 0.050 & 0.053 \\
Jitter $s$ [\kms] & 0.170 & 0.150 & 0.090 & 0.81 & 0.40 & 0.36 & 0.107 & 0.070 & 0.056 \\
\feh             & +0.325 & 0.045 & 0.042 & $-$0.318 & 0.043& 0.044 & +0.010 & 0.041 & 0.038 \\
$u_1$\,\tablenotemark{d}	 & 0.4714 & -- & -- & 0.3272 & -- & -- & 0.3921 & -- & -- \\
$u_2$\,\tablenotemark{d}	 & 0.2185 & -- & -- & 0.2971 & -- & -- & 0.2630 & -- & -- \\
$F_2/F_1$\,\tablenotemark{e} [ppm] & $<$190  & -- & -- & 560 & 160 & 180 & $<$97 & -- & -- \\
\hline

\multicolumn{9}{l}{\it Derived parameters}\\
$R_2$\ [\rsun]  & 0.1702 & 0.0046 & 0.0032 & 0.1996 & 0.0119 & 0.0067 & 0.0913 & 0.0048 & 0.0026\\
$K$\ [\kms]       & 12.53 & 0.10 & 0.10 & 12.00 & 0.14 & 0.14 & 8.720 & 0.069 & 0.074 \\
$a/R_1$       & 14.66 & 0.23 & 0.33 & 25.80 & 0.98 & 1.06 & 19.17 & 0.52 & 0.89 \\
\teff\ [K]      & 5579 & 77 & 78 & 5908 & 64 & 63 & 5747 & 70 & 64 \\
\logg\ [cgs]    & 4.01 & 0.011 & 0.015 & 4.104 & 0.024 & 0.038 & 4.288 & 0.033 & 0.020 \\
$L_1\,[L_\odot]$ & 2.52 & 0.39 & 0.22 & 2.17 & 0.31 & 0.22 & 1.36 & 0.13 & 0.12 \\
$L_2\,[L_\odot]$\,\tablenotemark{e} & $<0.00010$  & -- & -- & 0.00114 & 0.00037 & 0.00043 & $< 0.00011$ & -- & -- \\
 Age [Gyr]       & 8.49 & 0.97 & 1.35 & 10.4 &1.2 & 1.0 & 9.16 & 0.93 & 0.91 \\
$a$\ [au]       & 0.11545 &0.00091 & 0.00087 & 0.1687 & 0.0017 & 0.0017 & 0.10439 & 0.00093 & 0.00090\\
$b$             & 0.118 & 0.119 & 0.082 & 0.20 & 0.15 & 0.14 & 0.24 & 0.14 & 0.16 \\
$T_{14}$\ [d]   & 0.2222 & 0.0016 & 0.0014 & 0.2533 & 0.0031 & 0.0022 & 0.2380  & 0.0020& 0.0013\\
$T_{12}$\ [d]   & 0.02051 & 0.00102 & 0.00034 & 0.03276 & 0.0034 & 0.0013 & 0.0183 & 0.0021 & 0.0010 \\
$e$             & 0.3797 & 0.0058 & 0.0090 & 0.360 & 0.016 & 0.018 & 0.2545 & 0.0065 & 0.0070\\
$\omega$\ [deg] & 40.7 & 1.1 & 1.7 & 48.6 & 3.9 & 3.6 & $-$126.1& 1.5& 1.5\\
Occultation Phase\tablenotemark{f}  & 0.6579 & 0.0048 & 0.0076 & 0.6764 & 0.0015 & 0.0021 & 0.3692 & 0.0046 & 0.0047 \\
$b_\mathrm{occ}$ & 0.20 & 0.22 & 0.14 & 0.33 & 0.24 & 0.22 & 0.18 & 0.10 & 0.12 \\
\enddata

\tablenotetext{a}{Derived using \texttt{SpecMatch} analysis of the spectra.}
\tablenotetext{b}{Model fit parameters, fitted to the \kt\ light curve, RVs, and stellar isochrones. Gaussian priors are applied to \feh\ using the values derived from the \texttt{SpecMatch} spectroscopic analysis. See \secr{anal} for more information.}
\tablenotetext{c}{For \kb\ and \kc\ $\gamma$ is the binary system's center of mass RV since the HIRES RVs are on an absolute scale. For \ka\ the RV of the template spectrum ($-57.87\pm0.10$ \kms, see \secr{tres}) needs to be added to $\gamma$ to get the center of mass RV.}
\tablenotetext{d}{Parameter fixed during the model fitting process.}
\tablenotetext{e}{\sig{3} upper limit given when no eclipse was detected.}
\tablenotetext{f}{The transit, or primary eclipse, is at phase zero.}

\end{deluxetable}


\begin{thebibliography}{}

\bibitem[Akeson et al.(2013)]{akeson13} Akeson, R.~L., Chen, X., Ciardi, D., et al.\ 2013, \pasp, 125, 989 

\bibitem[Baraffe et al.(2003)]{baraffe03} Baraffe, I., Chabrier, G., Barman, T.~S., Allard, F., \& Hauschildt, P.~H.\ 2003, \aap, 402, 701 

\bibitem[Baraffe et al.(2015)]{baraffe15} Baraffe, I., Homeier, D., Allard, F., \& Chabrier, G.\ 2015, \aap, 577, A42 

\bibitem[Bayliss et al.(2017)]{bayliss17} Bayliss, D., Hojjatpanah, S., Santerne, A., et al.\ 2017, \aj, 153, 15 

\bibitem[Borucki(2016)]{borucki16} Borucki, W.~J.\ 2016, Reports on Progress in Physics, 79, 036901 

\bibitem[Buchhave et al.(2010)]{buchhave10} Buchhave, L.~A., Bakos, G.~{\'A}., Hartman, J.~D., et al.\ 2010, \apj, 720, 1118 

\bibitem[Burrows et al.(2011)]{borrows11} Burrows, A., Heng, K., \& Nampaisarn, T.\ 2011, \apj, 736, 47 

\bibitem[Cabrera et al.(2017)]{cabrera17} Cabrera, J., Barros, S.~C.~C., Armstrong, D., et al.\ 2017, arXiv:1707.08007 

\bibitem[Chubak et al.(2012)]{chubak12} Chubak, C., Marcy, G., Fischer, D.~A., et al.\ 2012, arXiv:1207.6212 

\bibitem[Claret(2004)]{claret04} Claret, A.\ 2004, \aap, 428, 1001 

\bibitem[Coughlin et al.(2016)]{coughlin16} Coughlin, J.~L., Mullally, F., Thompson, S.~E., et al.\ 2016, \apjs, 224, 12 

\bibitem[Crossfield et al.(2016)]{crossfield16} Crossfield, I.~J.~M., Ciardi, D.~R., Petigura, E.~A., et al.\ 2016, \apjs, 226, 7 

\bibitem[D{\'{\i}}az et al.(2013)]{diaz13} D{\'{\i}}az, R.~F., Damiani, C., Deleuil, M., et al.\ 2013, \aap, 551, L9 

\bibitem[D{\'{\i}}az et al.(2014)]{diaz14} D{\'{\i}}az, R.~F., Montagnier, G., Leconte, J., et al.\ 2014, \aap, 572, A109 

\bibitem[Dotter et al.(2008)]{dotter08} Dotter, A., Chaboyer, B., Jevremovi{\'c}, D., et al.\ 2008, \apjs, 178, 89-101 

\bibitem[Foreman-Mackey et al.(2013)]{emcee13} Foreman-Mackey, D., Hogg, D.~W., Lang, D., \& Goodman, J.\ 2013, \pasp, 125, 306 

\bibitem[Fressin et al.(2013)]{fressin13} Fressin, F., Torres, G., Charbonneau, D., et al.\ 2013, \apj, 766, 81 

\bibitem[{{F{\H u}r{\'e}sz}(2008)}]{furesz08} {F{\H u}r{\'e}sz}, G. 2008, PhD thesis, {Univ. of Szeged, Hungary}

\bibitem[Haywood et al.(2016)]{haywood16} Haywood, R.~D., Collier Cameron, A., Unruh, Y.~C., et al.\ 2016, \mnras, 457, 3637 

\bibitem[Howard et al.(2009)]{howard09} Howard, A.~W., Johnson, J.~A., Marcy, G.~W., et al.\ 2009, \apj, 696, 75 

\bibitem[Howell et al.(2014)]{howell14} Howell, S.~B., Sobeck, C., Haas, M., et al.\ 2014, \pasp, 126, 398 

\bibitem[Mandel \& Agol(2002)]{mandel02} Mandel, K., \& Agol, E.\ 2002, \apjl, 580, L171 

\bibitem[Mazeh(2008)]{mazeh08} Mazeh, T.\ 2008, EAS Publications Series, 29, 1 

\bibitem[Montet et al.(2015)]{montet15} Montet, B.~T., Morton, T.~D., Foreman-Mackey, D., et al.\ 2015, \apj, 809, 25 

\bibitem[Morton(2012)]{morton12} Morton, T.~D.\ 2012, \apj, 761, 6 

\bibitem[Morton(2015a)]{morton15a} Morton, T.~D.\ 2015a, Astrophysics Source Code Library, ascl:1503.010 

\bibitem[Morton(2015b)]{morton15b} Morton, T.~D.\ 2015b, Astrophysics Source Code Library, ascl:1503.011 

\bibitem[Morton et al.(2016)]{morton16} Morton, T.~D., Bryson, S.~T., Coughlin, J.~L., et al.\ 2016, \apj, 822, 86 

\bibitem[Moutou et al.(2013)]{moutou13} Moutou, C., Bonomo, A.~S., Bruno, G., et al.\ 2013, \aap, 558, L6 

\bibitem[Nidever et al.(2002)]{nidever02} Nidever, D.~L., Marcy, G.~W., Butler, R.~P., Fischer, D.~A., \& Vogt, S.~S.\ 2002, \apjs, 141, 503 

\bibitem[Ofir et al.(2012)]{ofir12} Ofir, A., Gandolfi, D., Buchhave, L., et al.\ 2012, \mnras, 423, L1 

\bibitem[Petigura(2015)]{petigura15} Petigura, E.~A.\ 2015, Ph.D.~Thesis, 

\bibitem[Petigura et al.(2017)]{petigura17} Petigura, E.~A., Howard, A.~W., Marcy, G.~W., et al.\ 2017, \aj, 154, 107 

\bibitem[Pont et al.(2005)]{pont05} Pont, F., Melo, C.~H.~F., Bouchy, F., et al.\ 2005, \aap, 433, L21 

\bibitem[Pont et al.(2006)]{pont06} Pont, F., Moutou, C., Bouchy, F., et al.\ 2006, \aap, 447, 1035 

\bibitem[Rauer et al.(2014)]{rauer14} Rauer, H., Catala, C., Aerts, C., et al.\ 2014, Experimental Astronomy, 38, 249 

\bibitem[Shporer et al.(2016)]{shporer16} Shporer, A., Fuller, J., Isaacson, H., et al.\ 2016, \apj, 829, 34 

\bibitem[Shporer et al.(2017)]{shporer17} Shporer, A., Zhou, G., Fulton, B.~J., et al.\ 2017, arXiv:1708.07128 

\bibitem[Southworth(2011)]{southworth11} Southworth, J.\ 2011, \mnras, 417, 2166 

\bibitem[Sozzetti et al.(2007)]{sozzetti17} Sozzetti, A., Torres, G., Charbonneau, D., et al.\ 2007, \apj, 664, 1190 

\bibitem[Sullivan et al.(2015)]{sullivan15} Sullivan, P.~W., Winn, J.~N., Berta-Thompson, Z.~K., et al.\ 2015, \apj, 809, 77 

\bibitem[Tal-Or et al.(2011)]{talor11} Tal-Or, L., Santerne, A., Mazeh, T., et al.\ 2011, \aap, 534, A67 

\bibitem[Torres et al.(2011)]{torres11} Torres, G., Fressin, F., Batalha, N.~M., et al.\ 2011, \apj, 727, 24 

\bibitem[Torres(2013)]{torres13} Torres, G.\ 2013, Astronomische Nachrichten, 334, 4

\bibitem[Torres et al.(2015)]{torres15} Torres, G., Kipping, D.~M., Fressin, F., et al.\ 2015, \apj, 800, 99 

\bibitem[Triaud et al.(2013)]{triaud13} Triaud, A.~H.~M.~J., Hebb, L., Anderson, D.~R., et al.\ 2013, \aap, 549, A18 

\bibitem[Vanderburg \& Johnson(2014)]{vanderburg14} Vanderburg, A., \& Johnson, J.~A.\ 2014, \pasp, 126, 948 

\bibitem[Vanderburg et al.(2016)]{vanderburg16} Vanderburg, A., Latham, D.~W., Buchhave, L.~A., et al.\ 2016, \apjs, 222, 14 

\bibitem[von Boetticher et al.(2017)]{vonboetticher17} von Boetticher, A., Triaud, A.~H.~M.~J., Queloz, D., et al.\ 2017, \aap, 604, L6 

\bibitem[Zhou et al.(2014)]{zhou14} Zhou, G., Bayliss, D., Hartman, J.~D., et al.\ 2014, \mnras, 437, 2831 

\end{thebibliography}
\end{document}